\documentclass[prb,showpacs,floatfix,twocolumn,byrevtex,groupedaddress]{revtex4-1}
\usepackage[english]{babel}
\usepackage{amsmath}
\usepackage{graphicx}
\usepackage{float}
\usepackage{bm}
\usepackage{multirow}
\usepackage{dcolumn}

\usepackage[dvipsnames,usenames]{color}
\begin{document}

\bibliographystyle{apsrev}

\title{Edge superconducting state in Nb thin film with rectangular arrays of antidots}
\author{W. J. Zhang, S. K. He, H. F. Liu, G. M. Xue, H. Xiao, B. H. Li, Z. C. Wen, X. F. Han, S. P. Zhao, C. Z. Gu, X. G. Qiu}
\email{xgqiu@iphy.ac.cn}
\affiliation{Beijing National Laboratory for Condensed Matter Physics, Institute of Physics, Chinese Academy of Sciences, P.O. Box 603, Beijing 100190, China}
\author{Victor V. Moshchalkov}
\affiliation{INPAC-Institute for Nanoscale Physics and Chemistry, KU Leuven, Celestijnenlaan 200D, B-3001 Leuven, Belgium}

%
\begin{abstract}
Superconducting Nb thin films with rectangular arrays of submicron antidots have been systemically investigated by transport measurements. In low fields, the magnetoresistance curves demonstrate well-defined dips at integral and rational numbers of flux quanta per unit cell, which corresponds to a superconducting wire network-like regime. When the magnetic field is higher than a saturation field, interstitial vortices interrupt the collective oscillation in low fields and form vortex sublattice, where a larger magnetic field interval is observed. In higher fields, a crossover behavior from the interstitial sublattice state to a single-loop-like state is observed, characterized by oscillations with a period of $\Phi_0/\pi r_{eff}^2$, originating from the existence of edge superconducting states with a size $r_{eff}$ around the antidots.
\end{abstract}
\maketitle

%
%
\section{INTRODUCTION}
Experiments on mesoscopic superconductors with dimensions comparable to the superconducting characteristic length scales have demonstrated that the sample topology strongly influences the superconducting properties, such as the phase boundary $T_c(H)$, magnetoresistance $R(H)$, and field dependent critical current $I_c(H)$.\cite{Little.PRL.1962,Moshchalkov.NATURE.1995,Berdiyorov.PRB.2009} Various topologies (single loops,\cite{Little.PRL.1962,Berdiyorov.PRB.2009} multiloops,\cite{Puig.PRB.1998} large infinite networks,\cite{Pannetier.PRL.1984} and arrays of antidots\cite{bezryadin.JLTP.1995,rosseel.PHYSICA.1997}) have been studied both experimentally and theoretically. 

In a perpendicular magnetic field, localized superconducting state can first nucleate near the edge of samples within a thin layer of width $W_s$ $\sim$ $\xi(T)$, similar to the nucleation of the surface superconductivity.\cite{saint.PL.1963} The so-called edge superconducting state\cite{bezryadin.JLTP.1996,Veauvy.PRB.2004} has an enhanced critical field $H_{c3}(T)$. The enhancement of $H_{c3}(T)$ above the bulk critical field $H_{c2}(T)$ greatly depends on the curvature of the superconducting/normal interface and the surface-to-volume ratio.\cite{bezryadin.JLTP.1995,moshchalkov.CONNECTIVITY.2000} Much higher enhancement of the ratio $H_{c3}/H_{c2}$ up to 3.6 has been observed in Pb thin film with a dense square antidot lattice.\cite{moshchalkov.CONNECTIVITY.2000}

For the square arrays of antidots, when the narrowest separation $\Delta W$ between neighboring holes is smaller than a critical value 1.84$\xi(T)$,\cite{saint.PL.1963} nucleation is dominated by the thin wire-like edge supercondcting states and the coupling between them. This kind of arrays is well described by the theory of supercondcting wire networks.\cite{Alexander.PRB.1983,Rammal.PRB.1983} The oscillations of $T_c(H)$ or $R(H)$ in such array are known as collective oscillations or network-like oscillations, whose period is corresponding to the area of the unit cell.\cite{Pannetier.PRL.1984} As shown by Bezryadin and Pannetier,\cite{bezryadin.JLTP.1995} when the magnetic field is high enough, a crossover behavior in the $T_c(H)$ curve from collective oscillations to single-loop-like (`single object') oscillations has been observed. The period of the single-loop-like oscillations is determined by the effective hole (antidot) radius $r_{eff}$ = $r_h + \frac{1}{2}W_s$, where $r_h$ is the hole radius. Due to the decrease of $W_s$ with increasing magnetic field, the single-loop-like oscillations are non-periodic. Although the dimensional crossover behaviors have been observed in square array of antidots,\cite{bezryadin.JLTP.1995,bezryadin.JLTP.1996,rosseel.PHYSICA.1997} few experiments on other symmetry of arrays are performed to check the universality. It is therefore interesting to study the rectangular arrays by introducing an anisotropy to the square arrays. We increase the length of one side of the rectangular unit cell, while keep the other one constant. In this way, we obtain a series of rectangular arrays of antidots characterized by two features: (1) separation $\Delta W$ of the short side $a$ is smaller than $\xi(T)$; (2) $\Delta W$ of the long side $b$ is larger than $4\xi(T)$ (i.e.$~$little overlap of the edge states in this direction). Furthermore, for perforated samples, Abrikosov (interstitial) vortices can appear in the wide superconducting strips between holes, which are observed by imaging techniques.\cite{bezryadin.JLTP.1996,Veauvy.PRB.2004,Karapetrov.PRL.2005,Kramer.PRL.2009} Thus, for the rectangular arrays of antidots, an extra interstitial vortex state will be involved as magnetic field increases, in contrast to the crossover behavior found in the square arrays of antidots.\cite{bezryadin.JLTP.1995,bezryadin.JLTP.1996,rosseel.PHYSICA.1997}


In this paper, we have performed detailed systematic transport measurements on the rectangular arrays of antidots with various geometric parameters. We observe clearly successive crossover behaviors from network-like state to interstitial vortex state, then to the single-loop-like state.
Hysteresis effect is observed for the interstitial vortex state. The crossover fields are found strongly dependent on temperature, the aspect ratio of the unit cell, and hole size.

%
%

\section{EXPERIMENT}
High quality Nb thin films with a thickness of about 100 nm were deposited on SiO$_2$ (300 nm)/Si substrates by magnetron sputtering. The Nb thin films have a critical temperature $T_{c}$ of 8.910 K and a superconducting transition width of about 21 mK (10\% - 90\%$R_{n}$ criterion, where $R_n$ is the normal state resistance at 9 K). For standard transport measurements, three four-probe microbridges were patterned on a single Nb substrate with ultraviolet photolithography and etched by reactive ion etching (RIE) in O$_{2}$ and  SF$_{6}$ plasmas. Each Nb microbridge had a 60 $\mu$m width for current flowing and a 60 $\mu$m distance between two voltage connects. The substrates were spun with polymethyl metacrylate (PMMA) resist layer and baked at 170 $^{\circ}$C for 1 minute. Designed patterns were written by electron beam lithography (EBL) on PMMA resist and developed for 40 seconds in MIBK:IPA (1:3) solution. Finally, the samples were etched by RIE and unexposed PMMA was removed in acetone. Rectangular arrays of circular antidots were obtained in the centers of the Nb microbridges. Note that one of the Nb microbridges was intentionally unexposed and used as a reference.

%
%

\begin{figure}[htb]
  \includegraphics[width=0.9\columnwidth]{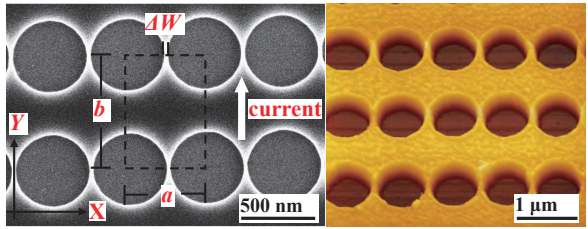}
  \caption{(color online) SEM and AFM images of a rectangular antidot lattice(sample S1), with a unit cell of $a$ $\times$ $b$ (800 nm $\times$ 1200 nm). The hole(antidot) radius is 375 nm and the distance between centers of holes is 800 nm, resulting in a width $\Delta W$ of the narrowest part of the constriction of 50 nm. The dashed rectangle indicates a unit cell.
}
  \label{Fig1}
\end{figure}

Figure$~$\ref{Fig1} shows images of scanning electron microscope (SEM) and atomic force microscope (AFM) for sample S1, with a rectangular unit cell of 800 nm $\times$ 1200 nm. The narrowest separation $\Delta W$ between neighboring holes along the X direction is 50 nm ($\pm$ 4 nm). The images demonstrate that the overall periodicity is maintained very well. Sharp edges are obtained after etching.

The measurements have been performed in Physical Properties Measurement System (PPMS-14, Quantum Design Inc.). Two phase lock-in amplifiers (SR830) are used for ac currents applied at a frequency of 30.9 Hz. The current is parallel to the long side of rectangular unit cell (Y direction). 
Applied magnetic field is perpendicular to the film surface. We sweep the magnetic field with a step of 0.4 Oe in low field regime and 1 Oe in high field regime. 
The temperature stability is better than 2 mK during the measurements. The superconducting coherence length $\xi(0)$ is 11.3 nm and the penetration depth $\lambda(0)$ is 74.0 nm, determined by measuring the $T_{c}(H)$ of the reference Nb microbridge.\cite{Tinkham.BOOK.1996} Thus, we have $\xi(t)$ = 113 nm, $\lambda_{eff}(t)$ $\approx$ $\lambda^2(t)/d$ = 1389.7 nm, at $t = 0.990$, where $t = T/T_c$ is the reduced temperature, $T$ $=$ 8.821 K, and $d$ = 100 nm is the thickness of thin film.

%
%
%
%

\begin{figure}[htb]
  \includegraphics[width=0.9\columnwidth]{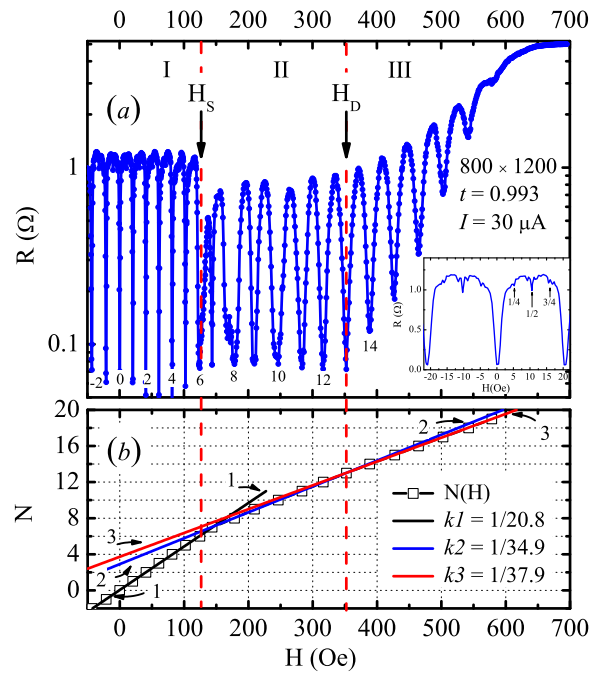}
  \caption{(color online) (a) Magnetoresistance of sample S1 is measured at $T$ = 8.640 K ($t = 0.993$) with $I$ = 30 $\mu$A. The dashed vertical lines divide $R(H)$ curve into three regions. The downward arrows indicate the positions of the crossover fields $H_S$ and $H_D$. Inset of (a): Enlarged plot of the low field regime. Fractional reduced magnetic fluxes are indicated with upward arrows. (b) Index of the $R(T)$ peaks $N$ as a function of the field position of the minima. Solid lines are linear fits for the data, with slopes ($k1$, $k2$ and $k3$), corresponding to the three regions in upper panel.
}
  \label{Fig2}
\end{figure}
%

%
%
\section{RESULTS AND DISCUSSION}
\subsection{Magnetoresistance and hysteresis effect}
The field dependent $R(H)$ curve for sample S1 (800 $\times$ 1200, $r_h$ = 375 nm) at $T$ = 8.640 K and current $I$ = 30 $\mu$A is given in Fig.$~$\ref{Fig2}(a). 
The $T_c$ of sample S1 is 8.702 K (50\%$R_{n}$ criterion). Three different regimes of the $R(H)$ curve can be distinguished by the shape of dips(minima) and the magnetic field interval $\Delta H$ between two consecutive dips: the low field (region-I), the intermediate field (region-II) and the high field (region-III) regimes. The downward arrows indicate the crossover fields at $H_S =$ 125 Oe and $H_D =$ 352 Oe. The number $N$ is a sequence number of dips relative to the one at zero field, which marks as $N$ = 0. We plot the number $N$ as a function of the position of the dips, as shown in Fig.$~$\ref{Fig2}(b). The data are fitted by straight lines with different slopes ($k1 = 1/20.8$, $k2 = 1/34.9$ and $k3 = 1/37.9$), corresponding to the three regions in the upper panel. The slope $k$ is a reciprocal of the average field interval $\overline{\Delta H}$. It is found that $\overline{\Delta H}$ in the three regions increases with field.

In the low field regime ($H < H_S$), $\overline{\Delta H}$ is 20.8 Oe, corresponding to one flux quantum per unit cell. It slightly deviates from the theoretic value $H_{R}$ = $\Phi_{0}/ab$ = 21.6 Oe, where $\Phi_{0}$ = 20.7 G-$\mu$m$ ^2$ is the flux quantum.\cite{Tinkham.BOOK.1996} The samples behave like the rectangular superconducting wire network, owing to the separations of neighboring holes comparable with $\xi(T)$ near $T_c$.\cite{Pannetier.PRL.1984,Moshchalkov.PRB.1998,Hoffmann.PRB.2000,Patel.PRB.2007} The inset of Fig.$~$\ref{Fig2}(a) shows a magnification of the low field $R(H)$ curve. The fractional reduced magnetic flux $f$ = $\Phi/\Phi_0$ = (1/4, 1/2, 3/4) is clearly visible. These features reflect the collective behavior of multiconnected superconducting wire network, 
which can be described by the mean field Ginzburg-Landau theory.\cite{bezryadin.JLTP.1995,bezryadin.JLTP.1996} The description of the collective behavior is different from the arrays with smaller and weaker pinning centers. In the latter case, the observed oscillatory dips in the magnetoresistance were explained by the vortex (multiquanta-vortex) matching model in the London limit.\cite{Reichhardt.PRL.1997,Martin.PRL.1999,Metlushko.PRB.1999,Reichhardt.PRB.2001,Stoll.PRB.2002,kemmler.PRL.2006}


The maxima of magnetoresistance show approximately the same magnitude in the low field regime, suggesting that the multiquanta vortex is effectively confined in each large antidot.\cite{bezryadin.JLTP.1996} When the magnetic field is larger than the saturation field $H_{S}$, the $R(H)$ behavior changes drastically. The collective oscillations are interrupted by the formation of additional vortices in the interstitial regions. The interstitial vortices cause phase decoherence events (i.e. phase slips\cite{Little.PR.1967}) or an incoherent `glassy' configuration of order parameter phase,\cite{itzler.PHYSICAB.1996} resulting in broad dips, missing of fine fractional structures, and increasing of field intervals. Similar phenomena have also been observed in rectangular arrays of magnetic dots, but explained by the reconfiguration transitions of vortex lattice.\cite{Martin.PRL.1999,Stoll.PRB.2002}

In the high field regime ($H > H_D$), a rapid increase of resistance background is observed and $\overline{\Delta H} = 37.9$ Oe is larger than that in region-II. As we will discuss below, the large period oscillations in region-III correspond to the single-loop-like behavior. Thus, for the single-loop-like oscillations, a radius $r_{0}$ deduced from $(\Phi_0/\pi\overline{\Delta H})^{1/2}$ $=$ 417 nm is comparable to the one given by the expression of $r_{eff}$ $=$ $(r_h + \Delta W) = 375 + 50 = 425$ nm. It should be mentioned that we have replaced $W_s$ by $\Delta W$, due to the fact that $\Delta W < \xi(T)$. Since $2r_{0} > a$, it demonstrates that the supercurrents strongly overlap in the thin wires along the Y direction.\cite{Velez.PRB.2002,Wordenweber.PRB.2004} To further study the $R(H)$ behavior, we performed hysteresis measurements at the same temperature and current. It is expected that a hysteresis effect occurs in the intermediate regime owing to the existence of interstitial vortices.

\begin{figure}[htb]
  \includegraphics[width=0.9\columnwidth]{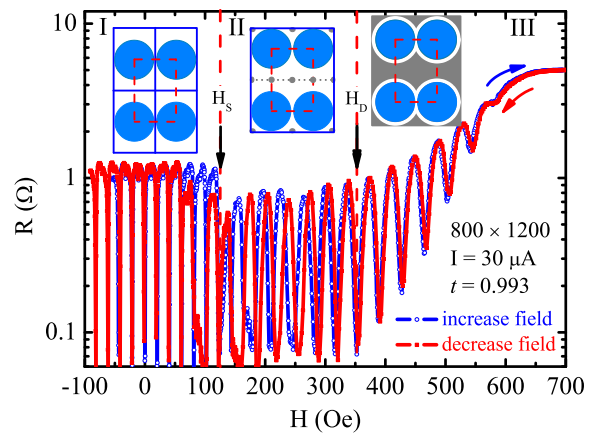}
  \caption{(color online) Hysteretic effect of the magnetoresistance for sample S1. Curves with open circles and filled squares correspond to increasing and decreasing field, respectively. The downward arrows mark the positions of $H_S$ and $H_D$, which divide the curves into three parts: (I) $H < H_S$, network-like region, where no vortices are located outside the holes; (II) $H_S < H <H_D$, interstitial vortex sublattice appears in the wide strips, with a normal core size of $2\xi(T)$; (III) $H > H_D$, the edge superconducting states are localized around each hole. The insets show schematic drawings of the vortex patterns for each region.
}
  \label{Fig3}
\end{figure}
%


In Fig.$~$\ref{Fig3}, hysteresis curves are recorded in the following way: the magnetic field first increases from -100 Oe to 750 Oe, then decreases from 750 Oe to -100 Oe. The curves are reversible in region I and III. In region-I, since only coreless vortices are present inside the holes, increasing or decreasing magnetic filed would produce the same magnetoresistance oscillations. 
In region-II, at first, every hole encloses 6$\Phi_0$. There is a competition between increasing the flux per hole and accommodating vortices at interstitial positions in the wide strips. The whole system favors to stay in the lowest free energy state. Thus, the additional vortices are expected to appear in the weak superconducting regions. Furthermore, the interstitial vortices in the wide strips can form stable sublattices, which produce dips in $R(H)$ curves.\cite{Metlushko.PRB.1999,Reichhardt.PRB.2001} The increase in $\overline{\Delta H}$ up to 34.9 Oe illustrates the presence of a compressed vortex distribution, compared to region-I. This increase is a little higher than the theoretical expectation for reconfiguration of square lattice $\Delta H_{Sq}$ = $\Phi_{0}/a^2$ = 32.3 $\pm$ 1.7 Oe.\cite{Martin.PRL.1999,Stoll.PRB.2002} The hysteresis effect in region-II is due to the intrinsic pinning forces and barrier for vortex motion coming from remaining superconducting regions.\cite{Harada.SCI.1996} With increasing field, the vortex patterns become more complex, and parts of the interstitial regions turn into normal state due to the penetration of the magnetic field. In region-III, most of the regions in S1 turns to normal states, except that edge states around holes are still superconducting. The curves become reversible again. A synchronized entrance of an additional vortex in each hole causes a resistance minimum in the $R(H)$, like the Little-Parks effect.\cite{Little.PRL.1962} 

From the hysteresis measurements, the boundaries of three regions can be clearly distinguished. Similar results have also been observed on sample S2 ($a$ = 800 nm, $b$ = 2000 nm, $r_h$ = 373 nm), with $H_{S}$ = 39 Oe and $H_{D}$ = 319 Oe at $T$ $=$ 8.640 K. In contrast to the work on magnetic dots,\cite{Stoll.PRB.2002} the hysteresis curves were measured just above or under the transition field, then the divisions of three regimes were possibly missed in their discussion.


\subsection{Parameters affecting the transitions}
\begin{figure}[htb]
  \includegraphics[width=0.9\columnwidth]{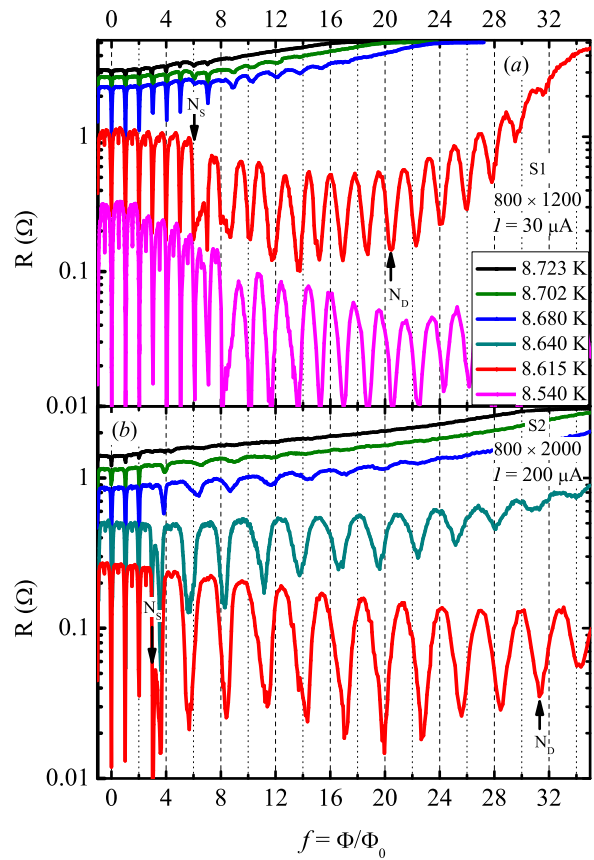}
  \caption{(color online) Resistance as a function of reduced magnetic flux $f$ for samples S1 [upper panel (a)] and S2 [lower panel (b)], measured at several temperatures with fixed currents $I$ = 30 $\mu$A and 200 $\mu$A, respectively. From top to bottom, the corresponding temperatures of $R(f)$ curves are 8.723 K, 8.702 K, 8.680 K, 8.615 K and 8.540 K (partly shown in the figures).
}
  \label{Fig4}
\end{figure}

Figure$~$\ref{Fig4} shows the $R(f)$ curves of samples S1 and S2 measured at several temperatures, with $I$ = 30 $\mu$A [panel (a)] and 200 $\mu$A [panel (b)], respectively. Since these two samples show very similar temperature dependence of the $R(f)$ curves, we focus our discussion on sample S1. In Fig.$~$\ref{Fig4}(a), from top to bottom, the temperature decreases from 8.723 K to 8.540 K (partly shown in the figures). In the network-like region of $R(f)$ curves, integer dips are always visible at these temperatures. Even some dips at fractional $f$ = (1/4, 1/3, 1/2) fields are well developed when the temperature is lower than 8.680 K. However, the magnetoresistance oscillations in the intermediate region and single-loop-like region are very sensitive to the temperature variation. At high temperatures close to $T_{c}$ ($T >$ 8.680 K), the oscillations in these two regions are broad and shallow. With decreasing temperature, the oscillations become more pronounced. With further decrease of the temperature, the oscillations become weaker and finally disappear. This illustrates different oscillation nature in these three regions.
%

The saturation number $N_S$ is defined by the largest possible number of the flux quanta trapped by an antidot. Mkrtchyan and Shmidt\cite{Mkrtchyan.JETP.1972} have theoretically estimated the maximum possible number of vortices trapped by a single insulating inclusion with an expression of $N_{St}(t)$ $=$ $r_h/2\xi(t)$. In our case, larger hole and smaller separation of antidots along the X direction are both used, where the edge superconductivity is more significant and produces effectively constraint to the flux.\cite{bezryadin.JLTP.1996,Veauvy.PRB.2004} As indicated by the arrow in Fig.$~$\ref{Fig4}(a), $N_S$ = 6 at $T$ = 8.615 K. It is larger than the theoretical value $N_{St}$($t$ = 0.99, $r_h$ $=$ 375 nm) $\sim$ 2. $N_S$ as a function of temperature for these two samples is shown in Fig.$~$\ref{Fig5}(a), compared with the theoretical value $N_{St}(t)$. As the temperature decreases, $N_S$ increases stepwisely. Interestingly, the experimental result for sample S2 is more close to the theoretical value at low temperature ($N_S$ $\approx$ 3, at $t$ = 0.978). This is due to the fact that the distributions of antidots in sample S2 are more sparse along the Y direction (or a larger aspect ratio $b/a$). In contrast, $N_S$ for sample S1 is nearly 2 $\sim$ 3 times of that for S2 at the same reduced temperature. In an array of antidots, the antidot-vortex interaction strongly affects the saturation number, resulting in a larger saturation number in the dense array.\cite{Doria.PHYSICAC.2000,Berdiyorov.PRB.2006} Besides that, the hole size,\cite{Moshchalkov.PRB.1998,Hoffmann.PRB.2000} and magnetic field\cite{Moshchalkov.PRB.1998} are also influencing $N_S$. The number of vortices inside a hole can still increase when the magnetic field further increases after the transition.\cite{Berdiyorov.PRB.2006} Therefore, the transition between region-I and region-II is mainly determined by the saturation number $N_S$, which is temperature, geometry and magnetic field dependent.

When the temperature is below 8.640 K in Fig.$~$\ref{Fig4}(a), we notice a decrease of resistance maxima in a wide range of magnetic field. This abnormal effect is related to the surface superconductivity around hole edges and it reaches its maximum near $H_{c2}$ (corresponding to $H_D$).\cite{fink.PRL.1965} Using this feature, we have plotted $H_{D}$ as a function of the reduced temperature, and obtained a linear temperature dependence, as shown in the inset of Fig.$~$\ref{Fig5}(b). It is found that $H_D(t)$ nearly coincides with the $H_{c2}(t)$ for the Nb thin film. 
We use the thin film expression $H^\ast_{c2}(t)$ = $\Phi_0/[2\pi\xi(t)^2]$ to roughly estimate the upper critical field. The theoretical value $H^\ast_{c2}(T/T_{c0} = 0.986)$ = 361 Oe is close to $H_{D}$ = 352 Oe, where $T$ $=$ 8.640 K, and $T_{c0}$ = 8.763 K is obtained from a linear extrapolation of line $H_D(t)$ to zero field. This suggests that $H_D$ can be indeed identified as $H_{c2}$ of sample S1. 
The region between $H_{c3}$ and $H_{c2}$ is the region where the bulk sample is already in the normal state and only a superconducting sheath persists at the surface of the sample parallel to the applied field. At a relatively low temperature, as the edge states merge with wide strips containing a finite value of order parameter, the whole array reentrances to the superconducting state. This has been found at $T$ = 8.540 K for S2. Another possible reason for the abnormal behavior is lowering mobility of interstitial vortices by a `caging effect'.\cite{Berdiyorov.PRB.2006}

By comparing the $R(H)$ curves of S1 with S2, we can study the influence of geometry (aspect ratio) on the magnetoresistance oscillation. However, there is no much difference in the position of the fine structures in the low field regime or the position of dips in the high field regime, except the difference in the saturation number $N_S$ and $H_D$. This implies that the oscillations in the $R(H)$ curves are mainly influenced by the conductivity of the arrays and upper critical fields. According to the above study, we find that the transition boundaries $H_S$ and $H_D$ are both temperature and geometry dependent. It is worth to note that the concept of rectangular array should hold in a certain range of aspect ratio (or length of $b$). Too large aspect ratio turns the system to one dimensional array of antidots, where the magnetoresistance behavior are dominated by the flux flow in the thin film.

\subsection{Discussion}
\begin{figure}[htb]
  \includegraphics[width=1\columnwidth]{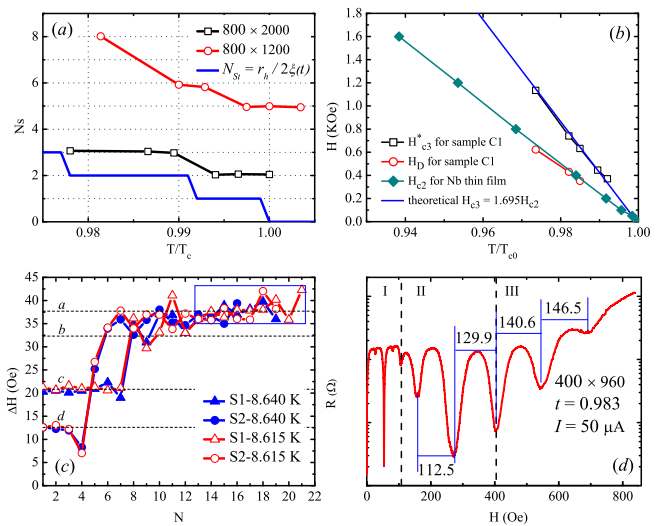}
  \caption{(color online) (a) Temperature dependence of the saturation numbers $N_S$ for sample S1 and S2, contrasted with the theoretical values $N_{St}(t)$ = $r_h/2\xi(t)$. (b) Phase boundary for arrays of antidots and a reference Nb thin film. $H_{c3}$ is calculated as $1.695H_{c2}$, where $H_{c2}$ is the experimental data for Nb thin film. $H_{c3}^\ast$ is the experimental data obtained from 90\%$R_n$ criterium in Fig.$~$\ref{Fig4}(a). (c) $\Delta H$ values as a function of index number $N$ for sample S1 ($\triangle$) and S2 ({\Large$\circ$}), are taken at $T$ = 8.640 K and 8.615 K, indicated by solid and open symbols, respectively. The horizontal dashed lines $a$ to $d$ mark the field values of 37.9, 32.3, 20.8 and 12.6 Oe. (d) $R(H)$ curve for the sample D1 with rectangular array of antidots (400 nm $\times$ 960 nm) measured at $t$ = 0.983 with $I$ = 50 $\mu$A.
}
  \label{Fig5}
\end{figure}

The well-defined shape of dips enables us to accurately determine $\Delta H$ between two consecutive dips. Figure$~$\ref{Fig5}(c) shows $\Delta H$ as a function of index number $N$ for samples S1 and S2, at $T$ = 8.640 K and 8.615 K. The horizontal dashed lines ($c$ and $d$) indicate $\overline{\Delta H}$ for samples S1 and S2 in region-I. We find that $\overline{\Delta H} = 20.8$ $\pm$ 1 Oe for S1, and $\overline{\Delta H}$ = 12.6 $\pm$ 0.9 Oe for C2. Two dashed lines ($a$ and $b$) mark the theoretical values $\Delta H_{SL}$ = $\Phi_{0}/(\pi r_{eff}^2)$ = 37.9 $\pm$ 2 Oe for the single-loop-like oscillations and $\Delta H_{Sq}$ = $\Phi_{0}/a^2$ =32.3 $\pm$ 1.7 Oe for square vortex lattice configuration.\cite{Martin.PRL.1999} The calculations are carried out with $r_{eff}$ = 417 $\pm$ 10 nm and a lattice constant $a$ = 800 $\pm$ 20 nm. It is found that most data are larger than the level of $\Delta H_{Sq}$ and fall into the region of $\Delta H_{SL}$ at higher field. The single-loop-like region marked with solid rectangle in Fig.$~$\ref{Fig5}(c) are full of data points larger than 35 Oe. The non-constant intervals are also consistent with the previous discussion. To further confirm our observations, samples with smaller unit cells are made to obtain a larger $\Delta H$ in region-III. The results for samples D1 (400 $\times$ 960, $r_h$ = 170 nm) are illustrated in Fig.$~$\ref{Fig5}(d). For D1, it is found that $N_S$ $=$ 2, which is smaller than that for S1, due to a smaller $r_h$. $\Delta H$ values in the range of 272 Oe $<$ H $<$ 402 Oe are close to the period of square lattice $H_{Sq}$= $\Phi_{0}/a^2$ = 129.4 $\pm$ 6 Oe. In the range of $H >$ 405 Oe, $\overline{\Delta H}$ is 143.6 Oe, which is close to the period calculated for the single-loop-like oscillations, with $r_{0}$ $=$ 216.5 nm $\sim$ $r_{eff}$ $=$ 230 nm. 
Similar results are also obtained in other samples (e.g. 400 $\times$ 655, 400 $\times$ 800 and 400 $\times$ 1000). Therefore, we have observed a series of crossover behaviors for the oscillations of magnetoresistance, like the samples S1 and S2.

Finally, we compare our results with previous works. Firstly, in contrast to the square arrays of antidots, an extra transition (interstitial vortex state) between collective oscillations and single-loop-like oscillations is found above the saturation field $H_S$. At higher fields, it indicates that the crossover from interstitial vortex state to the single-loop-like state is triggered by $H_{c2}$ of the sample, offering another way to induce crossover behaviors. Secondly, changes in the periodicity and the shape of the dips in magnetoresistance have been found in superconductors with magnetic dots, nonmagnetic dots and antidots.\cite{Martin.PRL.1999,Stoll.PRB.2002} The dominant mechanisms discussed in those works are analyzed in terms of two possible models: the reconfiguration model and the multivortex model. Following these pictures, our results require that vortices form a denser state in high field region. 
Whereas, there are difficulties with the explanation of the phenomena, such as the periodic appearances of the fine structures in region-I, the non-hysteresis effects in region-III and the much large field intervals (42 Oe $> \Delta H_{Sq}$, for S1) at fields higher than the upper critical field. On the other hand, from the viewpoint of dynamics of the vortex lattice ordering,\cite{Kokubo.PRL.2002} a monotonic increase in the intervals with increasing magnetic field is supposed to occur. However, this has not been found in our work. Thus, it seems to be inappropriate to use the London limit at such high temperatures and magnetic fields. We can discuss the results in the framework of the Ginzburg-Landau theory by considering the order parameter modulation. 
To clarify these effects, further direct imaging experiments, transport measurements and theoretical simulations in this type of arrays of antidots are necessary.

%
%

%
%
\section{Conclusion}
In conclusion, we have investigated magnetoresistance of superconducting Nb thin films containing rectangular arrays of large antidots. The $R(H)$ curve is divided into three regions by comparing the results with hysteresis measurements. At low magnetic fields, the system behaves like a weak link wire network, giving rise to dips in $R(H)$ at integral and fractional reduced magnetic flux below the saturation filed $H_S$. At the intermediate fields, the interstitial vortices form sublattice inducing larger magnetic field intervals in $R(H)$ and a hysteresis effect indicates the existence of the barrier for the vortex motion. As soon as the magnetic field exceeds $H_{c2}$, superconductivity nucleates near the edge of antidot in a layer with a width of $\xi(T)$, resulting in a single-loop-like superconducting state. In this state, the non-periodic oscillations and a fully reversible behavior are found in the $R(H)$ curves. The field intervals are determined from the effective diameter of hole, which is comparable with the short side constant of the unit cell. The crossover fields ($H_S$ and $H_D$) between the three regions are both temperature and geometry dependent.

%
%

\section*{Acknowledgement}
We thank Q. Niu and X. C. Xie for fruitful discussions. W. J. Zhang thanks for the help of microfabrication Lab and S. K. Su of group EX2 in IOP, CAS. This work is supported by National Basic Research Program of China (No.$~$2009CB929102, 2011CBA00107, 2012CB921302) and National Science Foundation of China (No.$~$10974241, 91121004). The work in at the KU Leuven is supported by the Methusalem Funding by the Flemish Government.
%
%

\begin{thebibliography}{35}
\expandafter\ifx\csname natexlab\endcsname\relax\def\natexlab#1{#1}\fi
\expandafter\ifx\csname bibnamefont\endcsname\relax
  \def\bibnamefont#1{#1}\fi
\expandafter\ifx\csname bibfnamefont\endcsname\relax
  \def\bibfnamefont#1{#1}\fi
\expandafter\ifx\csname citenamefont\endcsname\relax
  \def\citenamefont#1{#1}\fi
\expandafter\ifx\csname url\endcsname\relax
  \def\url#1{\texttt{#1}}\fi
\expandafter\ifx\csname urlprefix\endcsname\relax\def\urlprefix{URL }\fi
\providecommand{\bibinfo}[2]{#2}
\providecommand{\eprint}[2][]{\url{#2}}

\bibitem[{\citenamefont{Little and Parks}(1962)}]{Little.PRL.1962}
\bibinfo{author}{\bibfnamefont{W.~A.} \bibnamefont{Little}} \bibnamefont{and}
  \bibinfo{author}{\bibfnamefont{R.~D.} \bibnamefont{Parks}},
  \bibinfo{journal}{Phys. Rev. Lett.} \textbf{\bibinfo{volume}{9}},
  \bibinfo{pages}{9} (\bibinfo{year}{1962}).

\bibitem[{\citenamefont{Moshchalkov et~al.}(1995)\citenamefont{Moshchalkov,
  Gielen, Strunk, Jonckheere, Qiu, Van~Haesendonck, and
  Bruynseraede}}]{Moshchalkov.NATURE.1995}
\bibinfo{author}{\bibfnamefont{V.~V.} \bibnamefont{Moshchalkov}},
  \bibinfo{author}{\bibfnamefont{L.}~\bibnamefont{Gielen}},
  \bibinfo{author}{\bibfnamefont{C.}~\bibnamefont{Strunk}},
  \bibinfo{author}{\bibfnamefont{R.}~\bibnamefont{Jonckheere}},
  \bibinfo{author}{\bibfnamefont{X.}~\bibnamefont{Qiu}},
  \bibinfo{author}{\bibfnamefont{C.}~\bibnamefont{Van~Haesendonck}},
  \bibnamefont{and}
  \bibinfo{author}{\bibfnamefont{Y.}~\bibnamefont{Bruynseraede}},
  \bibinfo{journal}{Nature} \textbf{\bibinfo{volume}{373}},
  \bibinfo{pages}{319} (\bibinfo{year}{1995}).

\bibitem[{\citenamefont{Berdiyorov et~al.}(2009)\citenamefont{Berdiyorov, Yu,
  Xiao, Peeters, Hua, Imre, and Kwok}}]{Berdiyorov.PRB.2009}
\bibinfo{author}{\bibfnamefont{G.~R.} \bibnamefont{Berdiyorov}},
  \bibinfo{author}{\bibfnamefont{S.~H.} \bibnamefont{Yu}},
  \bibinfo{author}{\bibfnamefont{Z.~L.} \bibnamefont{Xiao}},
  \bibinfo{author}{\bibfnamefont{F.~M.} \bibnamefont{Peeters}},
  \bibinfo{author}{\bibfnamefont{J.}~\bibnamefont{Hua}},
  \bibinfo{author}{\bibfnamefont{A.}~\bibnamefont{Imre}}, \bibnamefont{and}
  \bibinfo{author}{\bibfnamefont{W.~K.} \bibnamefont{Kwok}},
  \bibinfo{journal}{Phys. Rev. B} \textbf{\bibinfo{volume}{80}},
  \bibinfo{pages}{064511} (\bibinfo{year}{2009}).

\bibitem[{\citenamefont{Puig et~al.}(1998)\citenamefont{Puig, Rosseel,
  Van~Look, Van~Bael, Moshchalkov, Bruynseraede, and
  Jonckheere}}]{Puig.PRB.1998}
\bibinfo{author}{\bibfnamefont{T.}~\bibnamefont{Puig}},
  \bibinfo{author}{\bibfnamefont{E.}~\bibnamefont{Rosseel}},
  \bibinfo{author}{\bibfnamefont{L.}~\bibnamefont{Van~Look}},
  \bibinfo{author}{\bibfnamefont{M.~J.} \bibnamefont{Van~Bael}},
  \bibinfo{author}{\bibfnamefont{V.~V.} \bibnamefont{Moshchalkov}},
  \bibinfo{author}{\bibfnamefont{Y.}~\bibnamefont{Bruynseraede}},
  \bibnamefont{and}
  \bibinfo{author}{\bibfnamefont{R.}~\bibnamefont{Jonckheere}},
  \bibinfo{journal}{Phys. Rev. B} \textbf{\bibinfo{volume}{58}},
  \bibinfo{pages}{5744} (\bibinfo{year}{1998}).

\bibitem[{\citenamefont{Pannetier et~al.}(1984)\citenamefont{Pannetier,
  Chaussy, Rammal, and Villegier}}]{Pannetier.PRL.1984}
\bibinfo{author}{\bibfnamefont{B.}~\bibnamefont{Pannetier}},
  \bibinfo{author}{\bibfnamefont{J.}~\bibnamefont{Chaussy}},
  \bibinfo{author}{\bibfnamefont{R.}~\bibnamefont{Rammal}}, \bibnamefont{and}
  \bibinfo{author}{\bibfnamefont{J.~C.} \bibnamefont{Villegier}},
  \bibinfo{journal}{Phys. Rev. Lett.} \textbf{\bibinfo{volume}{53}},
  \bibinfo{pages}{1845} (\bibinfo{year}{1984}).

\bibitem[{\citenamefont{Bezryadin and Pannetier}(1995)}]{bezryadin.JLTP.1995}
\bibinfo{author}{\bibfnamefont{A.}~\bibnamefont{Bezryadin}} \bibnamefont{and}
  \bibinfo{author}{\bibfnamefont{B.}~\bibnamefont{Pannetier}},
  \bibinfo{journal}{J. Low Temp. Phys.} \textbf{\bibinfo{volume}{98}},
  \bibinfo{pages}{251} (\bibinfo{year}{1995}).

\bibitem[{\citenamefont{Rosseel et~al.}(1997)\citenamefont{Rosseel, Puig,
  Baert, Van~Bael, Moshchalkov, and Bruynseraede}}]{rosseel.PHYSICA.1997}
\bibinfo{author}{\bibfnamefont{E.}~\bibnamefont{Rosseel}},
  \bibinfo{author}{\bibfnamefont{T.}~\bibnamefont{Puig}},
  \bibinfo{author}{\bibfnamefont{M.}~\bibnamefont{Baert}},
  \bibinfo{author}{\bibfnamefont{M.}~\bibnamefont{Van~Bael}},
  \bibinfo{author}{\bibfnamefont{V.}~\bibnamefont{Moshchalkov}},
  \bibnamefont{and}
  \bibinfo{author}{\bibfnamefont{Y.}~\bibnamefont{Bruynseraede}},
  \bibinfo{journal}{Physica C} \textbf{\bibinfo{volume}{282}},
  \bibinfo{pages}{1567} (\bibinfo{year}{1997}).

\bibitem[{\citenamefont{Saint-James and Gennes}(1963)}]{saint.PL.1963}
\bibinfo{author}{\bibfnamefont{D.}~\bibnamefont{Saint-James}} \bibnamefont{and}
  \bibinfo{author}{\bibfnamefont{P.}~\bibnamefont{Gennes}},
  \bibinfo{journal}{Phys. Letters} \textbf{\bibinfo{volume}{7}}
  (\bibinfo{year}{1963}).

\bibitem[{\citenamefont{Bezryadin and Pannetier}(1996)}]{bezryadin.JLTP.1996}
\bibinfo{author}{\bibfnamefont{A.}~\bibnamefont{Bezryadin}} \bibnamefont{and}
  \bibinfo{author}{\bibfnamefont{B.}~\bibnamefont{Pannetier}},
  \bibinfo{journal}{J. Low Temp. Phys.} \textbf{\bibinfo{volume}{102}},
  \bibinfo{pages}{73} (\bibinfo{year}{1996}).

\bibitem[{\citenamefont{Veauvy et~al.}(2004)\citenamefont{Veauvy, Hasselbach,
  and Mailly}}]{Veauvy.PRB.2004}
\bibinfo{author}{\bibfnamefont{C.}~\bibnamefont{Veauvy}},
  \bibinfo{author}{\bibfnamefont{K.}~\bibnamefont{Hasselbach}},
  \bibnamefont{and} \bibinfo{author}{\bibfnamefont{D.}~\bibnamefont{Mailly}},
  \bibinfo{journal}{Phys. Rev. B} \textbf{\bibinfo{volume}{70}},
  \bibinfo{pages}{214513} (\bibinfo{year}{2004}).

\bibitem[{\citenamefont{Moshchalkov et~al.}(2000)\citenamefont{Moshchalkov,
  Bruyndoncx, and Van~Look}}]{moshchalkov.CONNECTIVITY.2000}
\bibinfo{author}{\bibfnamefont{V.}~\bibnamefont{Moshchalkov}},
  \bibinfo{author}{\bibfnamefont{V.}~\bibnamefont{Bruyndoncx}},
  \bibnamefont{and} \bibinfo{author}{\bibfnamefont{L.}~\bibnamefont{Van~Look}},
  \bibinfo{journal}{Connectivity and superconductivity} pp.
  \bibinfo{pages}{87--137} (\bibinfo{year}{2000}).

\bibitem[{\citenamefont{Alexander}(1983)}]{Alexander.PRB.1983}
\bibinfo{author}{\bibfnamefont{S.}~\bibnamefont{Alexander}},
  \bibinfo{journal}{Phys. Rev. B} \textbf{\bibinfo{volume}{27}},
  \bibinfo{pages}{1541} (\bibinfo{year}{1983}).

\bibitem[{\citenamefont{Rammal et~al.}(1983)\citenamefont{Rammal, Lubensky, and
  Toulouse}}]{Rammal.PRB.1983}
\bibinfo{author}{\bibfnamefont{R.}~\bibnamefont{Rammal}},
  \bibinfo{author}{\bibfnamefont{T.~C.} \bibnamefont{Lubensky}},
  \bibnamefont{and} \bibinfo{author}{\bibfnamefont{G.}~\bibnamefont{Toulouse}},
  \bibinfo{journal}{Phys. Rev. B} \textbf{\bibinfo{volume}{27}},
  \bibinfo{pages}{2820} (\bibinfo{year}{1983}).

\bibitem[{\citenamefont{Karapetrov et~al.}(2005)\citenamefont{Karapetrov,
  Fedor, Iavarone, Rosenmann, and Kwok}}]{Karapetrov.PRL.2005}
\bibinfo{author}{\bibfnamefont{G.}~\bibnamefont{Karapetrov}},
  \bibinfo{author}{\bibfnamefont{J.}~\bibnamefont{Fedor}},
  \bibinfo{author}{\bibfnamefont{M.}~\bibnamefont{Iavarone}},
  \bibinfo{author}{\bibfnamefont{D.}~\bibnamefont{Rosenmann}},
  \bibnamefont{and} \bibinfo{author}{\bibfnamefont{W.~K.} \bibnamefont{Kwok}},
  \bibinfo{journal}{Phys. Rev. Lett.} \textbf{\bibinfo{volume}{95}},
  \bibinfo{pages}{167002} (\bibinfo{year}{2005}).

\bibitem[{\citenamefont{Kramer et~al.}(2009)\citenamefont{Kramer, Silhanek,
  Van~de Vondel, Raes, and Moshchalkov}}]{Kramer.PRL.2009}
\bibinfo{author}{\bibfnamefont{R.~B.~G.} \bibnamefont{Kramer}},
  \bibinfo{author}{\bibfnamefont{A.~V.} \bibnamefont{Silhanek}},
  \bibinfo{author}{\bibfnamefont{J.}~\bibnamefont{Van~de Vondel}},
  \bibinfo{author}{\bibfnamefont{B.}~\bibnamefont{Raes}}, \bibnamefont{and}
  \bibinfo{author}{\bibfnamefont{V.~V.} \bibnamefont{Moshchalkov}},
  \bibinfo{journal}{Phys. Rev. Lett.} \textbf{\bibinfo{volume}{103}},
  \bibinfo{pages}{067007} (\bibinfo{year}{2009}).

\bibitem[{\citenamefont{Tinkham}(1996)}]{Tinkham.BOOK.1996}
\bibinfo{author}{\bibfnamefont{M.}~\bibnamefont{Tinkham}},
  \emph{\bibinfo{title}{Introduction to Superconductivity, 2nd edition}}
  (\bibinfo{publisher}{McGraw-Hill}, \bibinfo{address}{New York},
  \bibinfo{year}{1996}).

\bibitem[{\citenamefont{Moshchalkov et~al.}(1998)\citenamefont{Moshchalkov,
  Baert, Metlushko, Rosseel, Van~Bael, Temst, Bruynseraede, and
  Jonckheere}}]{Moshchalkov.PRB.1998}
\bibinfo{author}{\bibfnamefont{V.~V.} \bibnamefont{Moshchalkov}},
  \bibinfo{author}{\bibfnamefont{M.}~\bibnamefont{Baert}},
  \bibinfo{author}{\bibfnamefont{V.~V.} \bibnamefont{Metlushko}},
  \bibinfo{author}{\bibfnamefont{E.}~\bibnamefont{Rosseel}},
  \bibinfo{author}{\bibfnamefont{M.~J.} \bibnamefont{Van~Bael}},
  \bibinfo{author}{\bibfnamefont{K.}~\bibnamefont{Temst}},
  \bibinfo{author}{\bibfnamefont{Y.}~\bibnamefont{Bruynseraede}},
  \bibnamefont{and}
  \bibinfo{author}{\bibfnamefont{R.}~\bibnamefont{Jonckheere}},
  \bibinfo{journal}{Phys. Rev. B} \textbf{\bibinfo{volume}{57}},
  \bibinfo{pages}{3615} (\bibinfo{year}{1998}).

\bibitem[{\citenamefont{Hoffmann et~al.}(2000)\citenamefont{Hoffmann, Prieto,
  and Schuller}}]{Hoffmann.PRB.2000}
\bibinfo{author}{\bibfnamefont{A.}~\bibnamefont{Hoffmann}},
  \bibinfo{author}{\bibfnamefont{P.}~\bibnamefont{Prieto}}, \bibnamefont{and}
  \bibinfo{author}{\bibfnamefont{I.~K.} \bibnamefont{Schuller}},
  \bibinfo{journal}{Phys. Rev. B} \textbf{\bibinfo{volume}{61}},
  \bibinfo{pages}{6958} (\bibinfo{year}{2000}).

\bibitem[{\citenamefont{Patel et~al.}(2007)\citenamefont{Patel, Xiao, Hua, Xu,
  Rosenmann, Novosad, Pearson, Welp, Kwok, and Crabtree}}]{Patel.PRB.2007}
\bibinfo{author}{\bibfnamefont{U.}~\bibnamefont{Patel}},
  \bibinfo{author}{\bibfnamefont{Z.~L.} \bibnamefont{Xiao}},
  \bibinfo{author}{\bibfnamefont{J.}~\bibnamefont{Hua}},
  \bibinfo{author}{\bibfnamefont{T.}~\bibnamefont{Xu}},
  \bibinfo{author}{\bibfnamefont{D.}~\bibnamefont{Rosenmann}},
  \bibinfo{author}{\bibfnamefont{V.}~\bibnamefont{Novosad}},
  \bibinfo{author}{\bibfnamefont{J.}~\bibnamefont{Pearson}},
  \bibinfo{author}{\bibfnamefont{U.}~\bibnamefont{Welp}},
  \bibinfo{author}{\bibfnamefont{W.~K.} \bibnamefont{Kwok}}, \bibnamefont{and}
  \bibinfo{author}{\bibfnamefont{G.~W.} \bibnamefont{Crabtree}},
  \bibinfo{journal}{Phys. Rev. B} \textbf{\bibinfo{volume}{76}},
  \bibinfo{pages}{020508} (\bibinfo{year}{2007}).

\bibitem[{\citenamefont{Reichhardt et~al.}(1997)\citenamefont{Reichhardt,
  Olson, and Nori}}]{Reichhardt.PRL.1997}
\bibinfo{author}{\bibfnamefont{C.}~\bibnamefont{Reichhardt}},
  \bibinfo{author}{\bibfnamefont{C.~J.} \bibnamefont{Olson}}, \bibnamefont{and}
  \bibinfo{author}{\bibfnamefont{F.}~\bibnamefont{Nori}},
  \bibinfo{journal}{Phys. Rev. Lett.} \textbf{\bibinfo{volume}{78}},
  \bibinfo{pages}{2648} (\bibinfo{year}{1997}).

\bibitem[{\citenamefont{Mart\'in et~al.}(1999)\citenamefont{Mart\'in, V\'elez,
  Hoffmann, Schuller, and Vicent}}]{Martin.PRL.1999}
\bibinfo{author}{\bibfnamefont{J.~I.} \bibnamefont{Mart\'in}},
  \bibinfo{author}{\bibfnamefont{M.}~\bibnamefont{V\'elez}},
  \bibinfo{author}{\bibfnamefont{A.}~\bibnamefont{Hoffmann}},
  \bibinfo{author}{\bibfnamefont{I.~K.} \bibnamefont{Schuller}},
  \bibnamefont{and} \bibinfo{author}{\bibfnamefont{J.~L.}
  \bibnamefont{Vicent}}, \bibinfo{journal}{Phys. Rev. Lett.}
  \textbf{\bibinfo{volume}{83}}, \bibinfo{pages}{1022} (\bibinfo{year}{1999}).

\bibitem[{\citenamefont{Metlushko et~al.}(1999)\citenamefont{Metlushko, Welp,
  Crabtree, Osgood, Bader, DeLong, Zhang, Brueck, Ilic, Chung
  et~al.}}]{Metlushko.PRB.1999}
\bibinfo{author}{\bibfnamefont{V.}~\bibnamefont{Metlushko}},
  \bibinfo{author}{\bibfnamefont{U.}~\bibnamefont{Welp}},
  \bibinfo{author}{\bibfnamefont{G.~W.} \bibnamefont{Crabtree}},
  \bibinfo{author}{\bibfnamefont{R.}~\bibnamefont{Osgood}},
  \bibinfo{author}{\bibfnamefont{S.~D.} \bibnamefont{Bader}},
  \bibinfo{author}{\bibfnamefont{L.~E.} \bibnamefont{DeLong}},
  \bibinfo{author}{\bibfnamefont{Z.}~\bibnamefont{Zhang}},
  \bibinfo{author}{\bibfnamefont{S.~R.~J.} \bibnamefont{Brueck}},
  \bibinfo{author}{\bibfnamefont{B.}~\bibnamefont{Ilic}},
  \bibinfo{author}{\bibfnamefont{K.}~\bibnamefont{Chung}},
  \bibnamefont{et~al.}, \bibinfo{journal}{Phys. Rev. B}
  \textbf{\bibinfo{volume}{60}}, \bibinfo{pages}{R12585}
  (\bibinfo{year}{1999}).

\bibitem[{\citenamefont{Reichhardt et~al.}(2001)\citenamefont{Reichhardt,
  Zim\'anyi, and Gr\o{}nbech-Jensen}}]{Reichhardt.PRB.2001}
\bibinfo{author}{\bibfnamefont{C.}~\bibnamefont{Reichhardt}},
  \bibinfo{author}{\bibfnamefont{G.~T.} \bibnamefont{Zim\'anyi}},
  \bibnamefont{and}
  \bibinfo{author}{\bibfnamefont{N.}~\bibnamefont{Gr\o{}nbech-Jensen}},
  \bibinfo{journal}{Phys. Rev. B} \textbf{\bibinfo{volume}{64}},
  \bibinfo{pages}{014501} (\bibinfo{year}{2001}).

\bibitem[{\citenamefont{Stoll et~al.}(2002)\citenamefont{Stoll, Montero,
  Guimpel, \AA{}kerman, and Schuller}}]{Stoll.PRB.2002}
\bibinfo{author}{\bibfnamefont{O.~M.} \bibnamefont{Stoll}},
  \bibinfo{author}{\bibfnamefont{M.~I.} \bibnamefont{Montero}},
  \bibinfo{author}{\bibfnamefont{J.}~\bibnamefont{Guimpel}},
  \bibinfo{author}{\bibfnamefont{J.~J.} \bibnamefont{\AA{}kerman}},
  \bibnamefont{and} \bibinfo{author}{\bibfnamefont{I.~K.}
  \bibnamefont{Schuller}}, \bibinfo{journal}{Phys. Rev. B}
  \textbf{\bibinfo{volume}{65}}, \bibinfo{pages}{104518}
  (\bibinfo{year}{2002}).

\bibitem[{\citenamefont{Kemmler et~al.}(2006)\citenamefont{Kemmler, G\"urlich,
  Sterck, P\"ohler, Neuhaus, Siegel, Kleiner, and Koelle}}]{kemmler.PRL.2006}
\bibinfo{author}{\bibfnamefont{M.}~\bibnamefont{Kemmler}},
  \bibinfo{author}{\bibfnamefont{C.}~\bibnamefont{G\"urlich}},
  \bibinfo{author}{\bibfnamefont{A.}~\bibnamefont{Sterck}},
  \bibinfo{author}{\bibfnamefont{H.}~\bibnamefont{P\"ohler}},
  \bibinfo{author}{\bibfnamefont{M.}~\bibnamefont{Neuhaus}},
  \bibinfo{author}{\bibfnamefont{M.}~\bibnamefont{Siegel}},
  \bibinfo{author}{\bibfnamefont{R.}~\bibnamefont{Kleiner}}, \bibnamefont{and}
  \bibinfo{author}{\bibfnamefont{D.}~\bibnamefont{Koelle}},
  \bibinfo{journal}{Phys. Rev. Lett.} \textbf{\bibinfo{volume}{97}},
  \bibinfo{pages}{147003} (\bibinfo{year}{2006}).

\bibitem[{\citenamefont{Little}(1967)}]{Little.PR.1967}
\bibinfo{author}{\bibfnamefont{W.~A.} \bibnamefont{Little}},
  \bibinfo{journal}{Phys. Rev.} \textbf{\bibinfo{volume}{156}},
  \bibinfo{pages}{396} (\bibinfo{year}{1967}).

\bibitem[{\citenamefont{Itzler and Chaikin}(1996)}]{itzler.PHYSICAB.1996}
\bibinfo{author}{\bibfnamefont{M.}~\bibnamefont{Itzler}} \bibnamefont{and}
  \bibinfo{author}{\bibfnamefont{P.}~\bibnamefont{Chaikin}},
  \bibinfo{journal}{Physica B} \textbf{\bibinfo{volume}{222}},
  \bibinfo{pages}{260} (\bibinfo{year}{1996}).

\bibitem[{\citenamefont{Velez et~al.}(2002)\citenamefont{Velez, Jaque,
  Mart\'in, Montero, Schuller, and Vicent}}]{Velez.PRB.2002}
\bibinfo{author}{\bibfnamefont{M.}~\bibnamefont{Velez}},
  \bibinfo{author}{\bibfnamefont{D.}~\bibnamefont{Jaque}},
  \bibinfo{author}{\bibfnamefont{J.~I.} \bibnamefont{Mart\'in}},
  \bibinfo{author}{\bibfnamefont{M.~I.} \bibnamefont{Montero}},
  \bibinfo{author}{\bibfnamefont{I.~K.} \bibnamefont{Schuller}},
  \bibnamefont{and} \bibinfo{author}{\bibfnamefont{J.~L.}
  \bibnamefont{Vicent}}, \bibinfo{journal}{Phys. Rev. B}
  \textbf{\bibinfo{volume}{65}}, \bibinfo{pages}{104511}
  (\bibinfo{year}{2002}).

\bibitem[{\citenamefont{W\"ordenweber et~al.}(2004)\citenamefont{W\"ordenweber,
  Dymashevski, and Misko}}]{Wordenweber.PRB.2004}
\bibinfo{author}{\bibfnamefont{R.}~\bibnamefont{W\"ordenweber}},
  \bibinfo{author}{\bibfnamefont{P.}~\bibnamefont{Dymashevski}},
  \bibnamefont{and} \bibinfo{author}{\bibfnamefont{V.~R.} \bibnamefont{Misko}},
  \bibinfo{journal}{Phys. Rev. B} \textbf{\bibinfo{volume}{69}},
  \bibinfo{pages}{184504} (\bibinfo{year}{2004}).

\bibitem[{\citenamefont{Harada et~al.}(1996)\citenamefont{Harada, Kamimura,
  Kasai, Matsuda, Tonomura, and Moshchalkov}}]{Harada.SCI.1996}
\bibinfo{author}{\bibfnamefont{K.}~\bibnamefont{Harada}},
  \bibinfo{author}{\bibfnamefont{O.}~\bibnamefont{Kamimura}},
  \bibinfo{author}{\bibfnamefont{H.}~\bibnamefont{Kasai}},
  \bibinfo{author}{\bibfnamefont{T.}~\bibnamefont{Matsuda}},
  \bibinfo{author}{\bibfnamefont{A.}~\bibnamefont{Tonomura}}, \bibnamefont{and}
  \bibinfo{author}{\bibfnamefont{V.~V.} \bibnamefont{Moshchalkov}},
  \bibinfo{journal}{Science} \textbf{\bibinfo{volume}{274}},
  \bibinfo{pages}{1167} (\bibinfo{year}{1996}).

\bibitem[{\citenamefont{Mkrtchyan and Shmidt}(1972)}]{Mkrtchyan.JETP.1972}
\bibinfo{author}{\bibfnamefont{G.}~\bibnamefont{Mkrtchyan}} \bibnamefont{and}
  \bibinfo{author}{\bibfnamefont{V.}~\bibnamefont{Shmidt}},
  \bibinfo{journal}{Sov. Phys. JETP} \textbf{\bibinfo{volume}{34}},
  \bibinfo{pages}{195} (\bibinfo{year}{1972}).

\bibitem[{\citenamefont{Doria et~al.}(2000)\citenamefont{Doria, de~Andrade, and
  Sardella}}]{Doria.PHYSICAC.2000}
\bibinfo{author}{\bibfnamefont{M.}~\bibnamefont{Doria}},
  \bibinfo{author}{\bibfnamefont{S.}~\bibnamefont{de~Andrade}},
  \bibnamefont{and} \bibinfo{author}{\bibfnamefont{E.}~\bibnamefont{Sardella}},
  \bibinfo{journal}{Physica C} \textbf{\bibinfo{volume}{341}},
  \bibinfo{pages}{1199} (\bibinfo{year}{2000}).


\bibitem[{\citenamefont{Berdiyorov et~al.}(2006)\citenamefont{Berdiyorov,
  Milo\v{s}evi\'{c}, and Peeters}}]{Berdiyorov.PRB.2006}
\bibinfo{author}{\bibfnamefont{G.~R.} \bibnamefont{Berdiyorov}},
  \bibinfo{author}{\bibfnamefont{M.~V.} \bibnamefont{Milo\v{s}evi\'{c}}},
  \bibnamefont{and} \bibinfo{author}{\bibfnamefont{F.~M.}
  \bibnamefont{Peeters}}, \bibinfo{journal}{Phys. Rev. B}
  \textbf{\bibinfo{volume}{74}}, \bibinfo{pages}{174512}
  (\bibinfo{year}{2006}).

\bibitem[{\citenamefont{Fink}(1965)}]{fink.PRL.1965}
\bibinfo{author}{\bibfnamefont{H.}~\bibnamefont{Fink}}, \bibinfo{journal}{Phys.
  Rev. Lett.} \textbf{\bibinfo{volume}{14}}, \bibinfo{pages}{309}
  (\bibinfo{year}{1965}).

\bibitem[{\citenamefont{Kokubo et~al.}(2002)\citenamefont{Kokubo, Besseling,
  Vinokur, and Kes}}]{Kokubo.PRL.2002}
\bibinfo{author}{\bibfnamefont{N.}~\bibnamefont{Kokubo}},
  \bibinfo{author}{\bibfnamefont{R.}~\bibnamefont{Besseling}},
  \bibinfo{author}{\bibfnamefont{V.~M.} \bibnamefont{Vinokur}},
  \bibnamefont{and} \bibinfo{author}{\bibfnamefont{P.~H.} \bibnamefont{Kes}},
  \bibinfo{journal}{Phys. Rev. Lett.} \textbf{\bibinfo{volume}{88}},
  \bibinfo{pages}{247004} (\bibinfo{year}{2002}).

\end{thebibliography}

\end{document}